\documentclass{webofc}
\usepackage[varg]{txfonts}   
\begin{document}
\title{Measurements of $p-\Lambda$ and $d-\Lambda$ correlations in 3 GeV Au+Au collisions at STAR}
%
%

\author{\firstname{Yu} \lastname{Hu}\inst{1}\fnsep\thanks{\email{yuhu@lbl.gov}} for the STAR collaboration
}

\institute{Lawrence Berkeley National Laboratory}

\abstract{%
Heavy-ion collisions provide a unique opportunity to explore nucleon-hyperon (N-Y) interactions through two-particle correlations. The $p-\Lambda$ and $d-\Lambda$ correlations shed light on both N-Y two-body and N-N-Y three-body interactions, which is crucial for understanding neutron star properties. We present the high precision measurement of $p-\Lambda$ and the first measurement of $d-\Lambda$ correlation with $\sqrt{s_{_{\rm NN}}}=$ 3 GeV Au+Au collisions at STAR. Using the Lednicky-Lyuboshitz formalism, we characterized emission source size, the scattering length ($f_0$), and the effective range ($d_0$) of $p-\Lambda$ and $d-\Lambda$ interactions. Using the $f_0$ and $d_0$ extracted from two spin states in $d-\Lambda$ correlation, the parameters from the doublet state indicate the hypertriton binding energy is consistent with the current average of world measurements. 
}

\maketitle

\section{Introduction}
 While the nuclear force has been studied for many decades through classical scattering experiments, measuring the interactions between nucleon (N) and hyperon (Y) is found to be very challenging. 
 Heavy-ion collisions is a well-designed laboratory to study the properties of the dense QCD matter. It offers us a new way to understand N-Y interactions by studying the two-particle correlation produced in a collision~\cite{Fabbietti:2020bfg}. 
 The correlations reveal valuable information about the space-time evolution of the particle-emitting source and final state interactions involving hyperons. That's the primary observable of interest in this proceedings. 

The momentum correlation between two identified particles can be presented as follows,
\begin{equation}
    C(\mathbf{p_{1}}, \mathbf{p_{2}}) \equiv \frac{P(\mathbf{p_1},\mathbf{p_2})}{P(\mathbf{p_1})\cdot P(\mathbf{p_2})},
    \label{cf_statistical}
\end{equation}
where the $\mathbf{p_1}$ and $\mathbf{p_2}$ are the momentum of particle 1 and 2. $P$ is the probability of finding such particle or particle pair with that certain momentum. As for the experimental measurement, we used the following equation to get the correlation function, which is equivalent to Eq~\ref{cf_statistical}.
\begin{equation}
    C(\mathbf{k^*}) = \mathcal{N}\frac{A(\mathbf{k^*})}{B(\mathbf{k^*})} = \int d^3 r^* S(\mathbf{r^*}) |\Psi(\mathbf{r^*}, \mathbf{k^*})|^2.
    \label{cf_experimental}
\end{equation}
Here the $A(\mathbf{k^*})$ is the distribution of $\mathbf{k^*}$ with both particles from the same event, $B(\mathbf{k^*})$ is for two particles from different events, and $\mathcal{N}$ is the normalization factor. $\mathbf{k^*}$ is the particle momentum in the pair rest frame. The correlation function can be expanded into the later part of Eq.~\ref{cf_experimental}. Here the $S(\mathbf{r^*})$ is the distribution of the relative distance of particle pairs, the $\Psi(\mathbf{r^*}, \mathbf{k^*})$ is the relative wave function of the particle pairs. 

If we take a smoothness approximation for the source function and expand the wave function in the Eq.~\ref{cf_experimental}. We can get
\begin{equation}
    C(\mathbf{k^{*}}) \approx 1+\frac{|f(k)|^2}{2r_{0}^2}F(d_{0})+\frac{2{\rm Re}f(k)}{\sqrt{\pi}r_0}F_{1}(2kr_0)-\frac{{\rm Im}f(k)}{r_{0}}F_{2}(2kr_0),
    \label{cf_LL}
\end{equation}
where,
\begin{equation}
    \frac{1}{f(k)} \approx \frac{1}{f_{0}}+\frac{d_{0}k^2}{2}-ik
    \label{cf_f0}.
\end{equation}
Here the $r_0$ is the radius parameter that defines the size of the source. In this study, the equivalent spherical Gaussian source radius of the particle pairs ($R_G$), which corresponds to the size of the emission source and $r_0$, is used to compare the size between two correlation systems. $f_0$ and $d_0$ are the scattering length and effective range, which are the parameters to describe the interaction between two particles. $F$, $F_1$, and $F_2$ are the parametric equations~\cite{Haidenbauer:2020uew}. This approach is called the Ledncky-Lyuboshitz (L-L) approach~\cite{Lednicky:1981su}.  With this approach, we can characterize the source radius and final state interaction from the measured correlation function. For the correlation between two nonidentical particles, the system can have different spin states. The $p-\Lambda$ systems have singlet (S) and triplet (T) spin states~\cite{Wang:1999bf}, and the $d-\Lambda$ systems have doublet (D) and quartet (Q) spin states~\cite{Haidenbauer:2020uew}. The $f_0$ and $d_0$ can be different in two different systems. The effective $f_0$ and $d_0$ are also widely used to describe the correlations when the statistics cannot separate two different spin states. 

\section{Correlation function and discussions}

The STAR (Solenoidal Tracker at RHIC) detector at RHIC (Relativistic Heavy Ion Collider) has collected many collision energies and systems over the two decades. For this study, we analyzed about 250M Au+Au collision events at center of mass energies per nucleon pair,$\sqrt{\rm s_{NN}}=3$ GeV which were collected in 2018 under the Fixed Target setup. 
We use the Time Projection Chamber (TPC) and the Time of Flight (ToF) detector to identify the pions ($\pi$), protons ($p$), and deuterons ($d$). We reconstructed the $\Lambda$ with the KFParticle package~\cite{Ju:2023xvg} using the $\pi$ and $p$. 

\begin{figure}[ht]
    \centering
    \includegraphics[width=\textwidth]{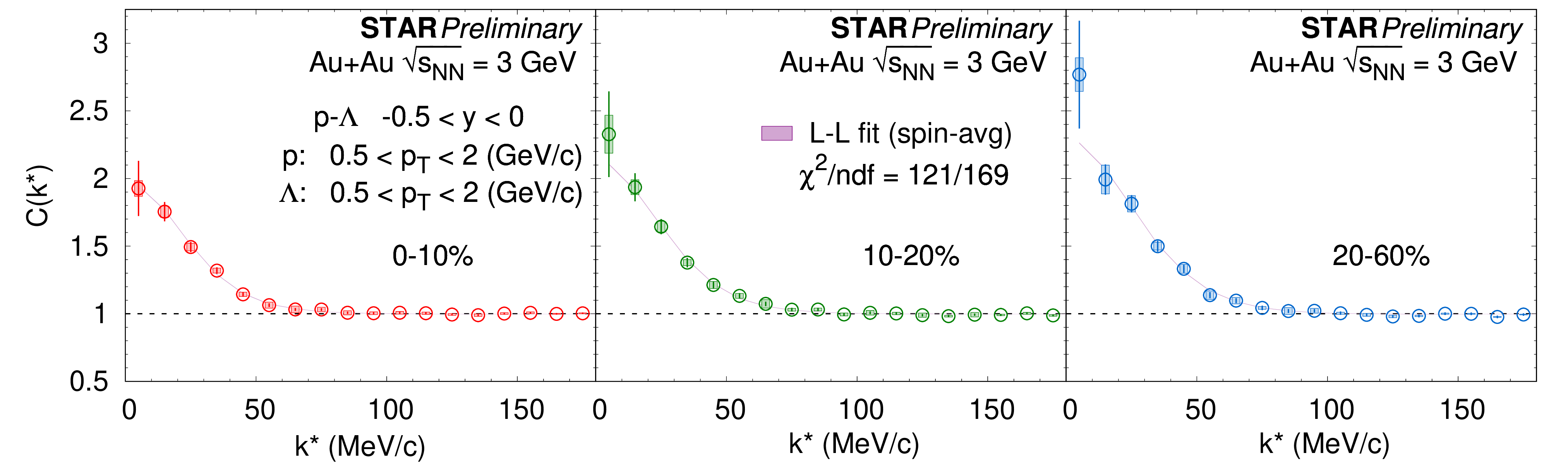}
    \caption{$p-\Lambda$ correlation measured in $-0.5<{\rm y}<0$, and $0.5<{p_T}<2$ GeV. The purple band shows the spin-averaged fitting results.}
    \label{cf_p_L}
\end{figure}

For the $p-\Lambda$ correlation measurement, we selected the $p_T$ from 0.5 GeV/c to 2 GeV/c for both $p$ and reconstructed $\Lambda$. We measured the correlation function in three rapidity ranges ($-0.5$ to 0, $-0.75$ to $-0.5$, and $-1$ to $-0.75$), and three different centralities (0-10\%, 10-20\%, and 20-60\%). The measured functions are corrected based on particle purity, feed-down contributions, track splitting effect, track merging effect, and momentum smearing effect. The examples of corrected correlation functions in $-0.5<{\rm y}<0$ for different centralities are shown in Figure~\ref{cf_p_L}. We applied a simultaneous fit to data in three different centralities and three different rapidities. The $f_0$ and $d_0$ are assumed to be the same for different centralities and rapidities, while the $R_G$ are assumed to be different for different centralities and rapidities in the fitting. We got the spin-averaged $f_0 = 2.32^{+0.12}_{-0.11}$ fm and $d_0 = 3.5^{+2.7}_{-1.3}$ fm. 

\begin{figure}[ht]
    \centering
    \includegraphics[width=\textwidth]{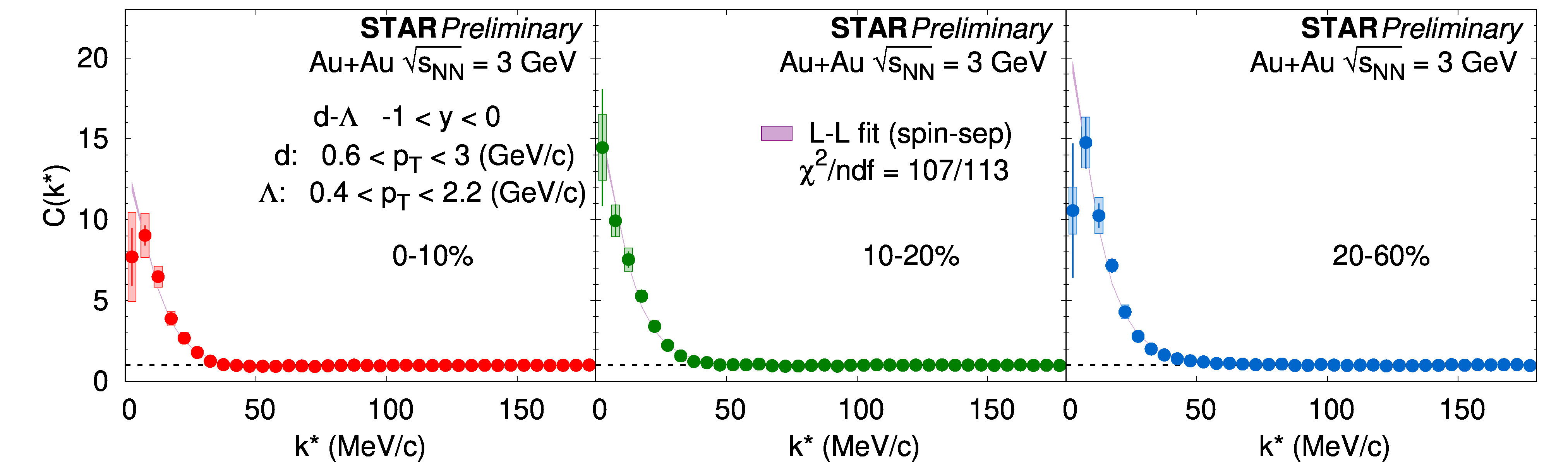}
    \caption{$d-\Lambda$ correlation measured in $-1<{\rm y}<0$. The purple band shows the spin-separated fit for both D and Q states. }
    \label{cf_d_L}
\end{figure}

For the $d-\Lambda$ correlation measurement, we used the $d$ selected in 0.6 GeV/c to 3 GeV/c, $\Lambda$ in 0.4 GeV/c to 2.2 GeV/c for the ${p_T}$, and from -1 to 0 for the rapidity range. The measurements are corrected for particle purity, track splitting, and merging effect. The $^3_{\Lambda}H$ contamination ($^3_{\Lambda}H \rightarrow d+p+\pi^-$) is also subtracted due to the indistinguishable experimental $p+\pi^-$ contribution to the $\Lambda$ reconstruction. Evaluation with the measured $^3_{\Lambda}H$ yields show $4\sim8\%$ of the $d-\Lambda$ entries come from $^3_{\Lambda}H$ decay at $k^*<100$ MeV/c in 10-20\% centrality~\cite{Kamada:1997rv}. The corrected  $d-\Lambda$ correlation functions are shown in Figure~\ref{cf_d_L}. A simultaneous fit to data in different centralities is shown in purple color. The fitting indicates $f_0(D)= -20^{+3}_{-3}$ fm, $d_0(D)= 3^{+2}_{-1}$ fm, $f_0(Q)= 16^{+2}_{-1}$ fm, and $d_0(Q)= 2^{+1}_{-1}$ fm.

\begin{figure}[ht]
    \centering
    \includegraphics[width=0.45\textwidth]{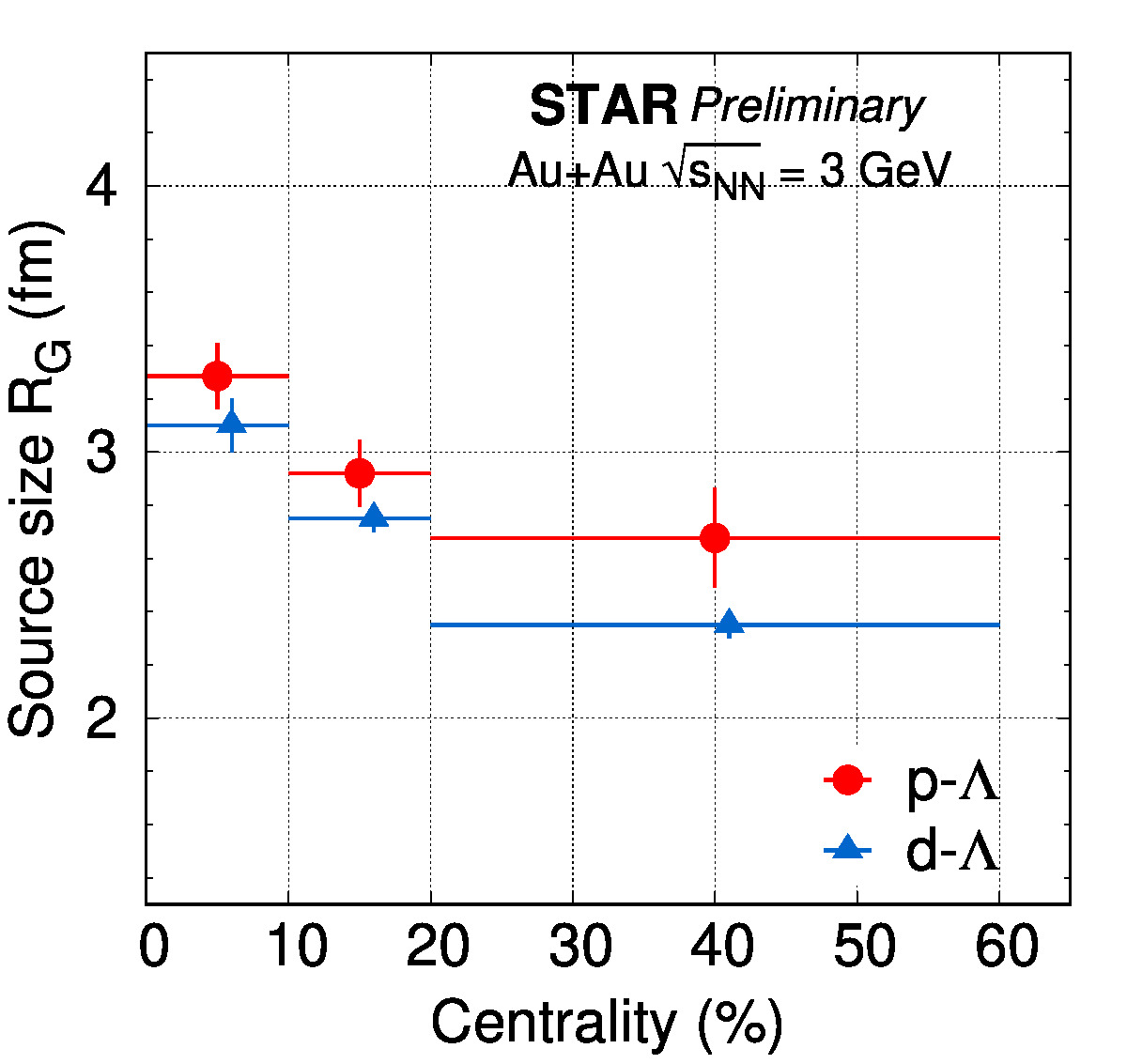}
    \includegraphics[width=0.44\textwidth]{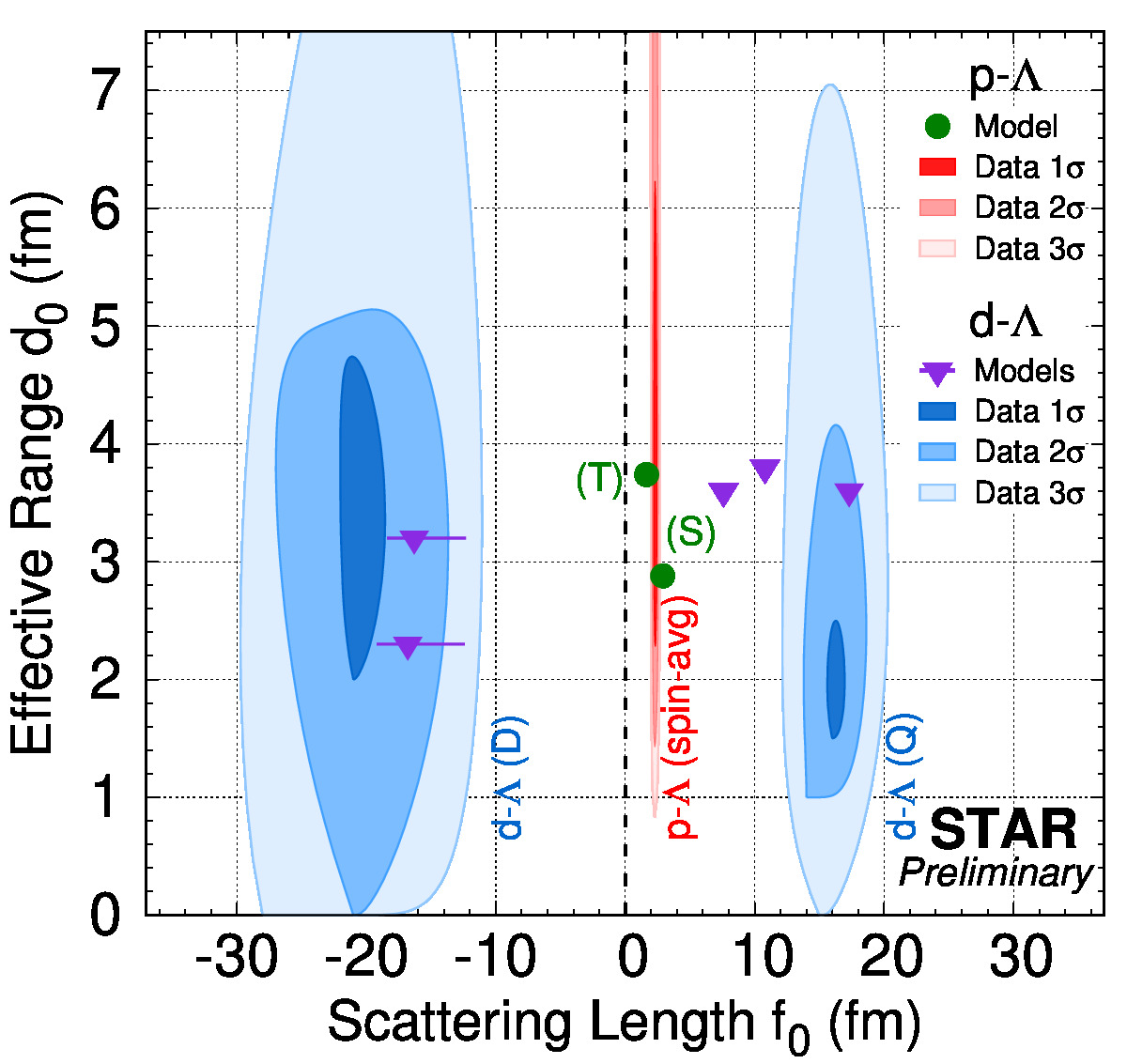}
    \caption{Left: Spherical Gaussian source of pairs ($R_G$); Right: scattering length ($f_0$) and effective range ($d_0$) with $1\sigma$, $2\sigma$, and $3\sigma$ from $p-\Lambda$ and $d-\Lambda$ correlations. The parameters are estimated using the Ledncky-Lyuboshitz (L-L) approach. Bezier smooth is applied to the right side contour plot for $d-\Lambda$ correlation to improve the visibility.}
    \label{fig_LL_fit_contour}
\end{figure}

The left plot of Figure~\ref{fig_LL_fit_contour} shows the $R_G$ extracted from the correlation functions. We successfully separated the emission source from the final state interaction. The observed $R_G$ is larger in more central collisions than the peripheral collisions. We also observed at a certain centrality, the $R_G$ for $d-\Lambda$ correlation is slightly smaller than for the $p-\Lambda$ correlation. These observations generally agree with our naive collision dynamics pictures that the central collision creates a bigger source, and if all the particles are produced at the same moment, the heavier particles fly slower so that the corresponding equivalent source is smaller for $d-\Lambda$.

The right-side plot of Figure~\ref{fig_LL_fit_contour} shows the $1\sigma$, $2\sigma$, and $3\sigma$ contour of the $f_0$ and $d_0$ for both systems. The green solid points show the theory predictions of the $f_0$ and $d_0$ for S and T states in $p-\Lambda$~\cite{Wang:1999bf}, and the purple points show the theory predicted D and Q states for $d-\Lambda$~\cite{Haidenbauer:2020uew}. The constraint of the $d_0$ is weaker because it characterizes the second order effect as shown in Eq.~\ref{cf_f0}. The larger magnitude $f_0$ in $d-\Lambda$ indicates a much stronger correlation and larger cross-section than the $p-\Lambda$. The negative $f_0 (D)$ indicates a bound state for $d-\Lambda$ systems which is the $^3_{\Lambda}{\rm H}$. 

Using $f_0 (D)$, $d_0 (D)$, and the Bethe formula from Effective Range Expansion (ERE), we could estimate the $^3_{\Lambda}{\rm H}$ binding energy. We got $^3_{\Lambda}{\rm H}~B_{\Lambda} = [0.04,0.33]$ (MeV) at 95\% CL. The result is consistent with the current world average~\cite{Chen:2023mel}. This measurement provides us with a new way to constrain the $^3_{\Lambda}{\rm H}$ structure. 

\section{Summary}

The $p-\Lambda$ correlation function is measured using the $\sqrt{s_{_{\rm NN}}}=$ 3 GeV Au+Au collisions from RHIC Beam Energy Scan II. 
The $d-\Lambda$ correlation is measured for the first time in the experiment. 
The emission source is successfully separated from the final state interactions. The $R_{G}$ is larger in central than peripheral collision, and slightly smaller in $d-\Lambda$ correlation than $p-\Lambda$ correlation.
The spin-averaged $f_0$ and $d_0$ is extracted to be $2.32^{+0.12}_{-0.11}$ fm and $3.5^{+2.7}_{-1.3}$ fm in $p-\Lambda$ correlation.  The effective fit showed $f_0(D)= -20^{+3}_{-3}$ fm, $d_0(D)= 3^{+2}_{-1}$ fm, $f_0(Q)= 16^{+2}_{-1}$ fm, and $d_0(Q)= 2^{+1}_{-1}$ fm in $d-\Lambda$ correlation. This measurement indicates the  $^3_{\Lambda}{\rm H}$ binding energy to be 0.04 to 0.33 MeV at 95\% CL, which provides a new method to study the hypernuclei structure in the heavy-ion collision experiment.

\bibliography{main}

\begin{thebibliography}{7}

\bibitem{Fabbietti:2020bfg}
L.~Fabbietti, V.~Mantovani~Sarti, O.~Vazquez~Doce, Ann. Rev. Nucl. Part. Sci. \textbf{71}, 377 (2021), \texttt{2012.09806}

\bibitem{Haidenbauer:2020uew}
J.~Haidenbauer, Phys. Rev. C \textbf{102}, 034001 (2020), \texttt{2005.05012}

\bibitem{Lednicky:1981su}
R.~Lednicky, V.L. Lyuboshits, Yad. Fiz. \textbf{35}, 1316 (1981)

\bibitem{Wang:1999bf}
F.q. Wang, S.~Pratt, Phys. Rev. Lett. \textbf{83}, 3138 (1999), \texttt{nucl-th/9907019}

\bibitem{Ju:2023xvg}
X.Y. Ju et~al., Nucl. Sci. Tech. \textbf{34}, 158 (2023)

\bibitem{Kamada:1997rv}
H.~Kamada, J.~Golak, K.~Miyagawa, H.~Witala, W.~Gloeckle, Phys. Rev. C \textbf{57}, 1595 (1998), \texttt{nucl-th/9709035}

\bibitem{Chen:2023mel}
J.~Chen, X.~Dong, Y.G. Ma, Z.~Xu (2023), \texttt{2311.09877}

\end{thebibliography}

\end{document}